# Reduced-order electrochemical models with shape functions for fast, accurate prediction of lithium-ion batteries under high C rates

Tianhan Gao[1] and Wei Lu[1,2,*]

[1] Department of Mechanical Engineering, University of Michigan, Ann Arbor, MI 48109, United States

[2] Department of Materials Science and Engineering, University of Michigan, Ann Arbor, MI 48109, United States

*e-mail: weilu@umich.edu

## Abstract

This paper proposes physical-based, reduced-order electrochemical models that are much faster than the electrochemical pseudo 2D (P2D) model, while providing high accuracy even under the challenging conditions of high C-rate and strong polarization of lithium ion concentration and potential in a battery cell. In particular, an innovative weak form of equations are developed by using shape functions, which reduces the fully coupled electrochemical and transport equations to ordinary differential equations, and provides self-consistent solutions for the evolution of the polynomial coefficients. Results show that the models, named as revised single-particle model (RSPM) and fast-calculating P2D model (FCP2D), give highly reliable prediction of battery operations, including under dynamic driving profiles. They can calculate battery parameters, such as terminal voltage, over-potential, interfacial current density, lithium-ion concentration distribution, and electrolyte potential distribution with a relative error less than 2%. Applicable for moderately high C rates (below 2.5 C), the RSPM is up to more than 33 times faster than the P2D model. The FCP2D is applicable for high C rates (above 2.5 C) and is about 8 times faster than the P2D model. With their high speed and accuracy, these physics-based models can significantly improve the capability and performance of the battery management system and accelerate battery design optimization.

**Keywords:** Lithium-ion battery; reduced-order model; revised single-particle-model (RSPM); fast-calculating P2D model (FCP2D); accuracy; efficiency

# 1. Introduction

The battery management system (BMS) is becoming increasingly important with the broad application of lithium-ion batteries in electric vehicles and consumer electronics [1,2]. The crucial foundation of a BMS is its underlying model to capture the battery behaviors, which determines the accuracy and efficiency for estimating and predicting the battery states, such state-of-charge (SOC) and state-of-health (SOH). The model directly affects the reliability and response speed of control strategies building on top of it, and the overall capability of a BMS to ensure cell and package performance, such as safety and energy/capacity utilization efficiency [3–5]. Current methods for battery modeling mainly include equivalent circuit model, electrochemical-based model, and data-driven model [6–8].

The equivalent circuit model (ECM) is a typical empirical model, which uses the elements of an electronic circuit (such as power source, resistor, capacitor and inductor) to represent the behavior of a battery cell. The ECM has been widely applied to predict the battery parameters, including terminal voltage, SOC, SOH, and electrochemical impedance spectroscopy (EIS) [9–11]. The ECM has the advantage of fast calculation since it does not need to solve the complex, coupled partial differential equations in electrochemical models [12]. However, the ECM cannot capture the underlying reaction kinetics, transport processes and intrinsic complex behaviors of a cell. The data-driven model is another type of empirical model, which uses a dataset to train a machine learning code to learn a relationship. Currently, data-driven models have been used to predict several battery parameters such as SOC, SOH and capacity [13]. Similar to the ECM, data-driven models also have challenges in describing the underlying physics in a battery cell. In addition to the demand of a large amount of training data and cost in initial training, data-driven models need re-training when there is a change of the cell chemistry or the operational condition.

The physics-based electrochemical model can accurately capture the underlying physical processes in a battery cell. The pseudo 2D (P2D) model originally proposed by Newman, Doyle, and Fuller [14,15] is widely used nowadays. However, a significant challenge of applying the P2D model is that many partial differential equations need to be solved simultaneously. It is computationally expensive and slow, which cannot be used directly in a BMS [16]. The single-particle model (SPM) [17] is a commonly used reduced-order model which simplifies the P2D model by approximating the two electrodes as two particles. The lithium-ion concentration in the electrolyte is assumed uniform, while the solid and electrolyte potentials in each electrode region are also assumed uniform. These simplifications make it possible to obtain an analytical expression for the cell terminal voltage. The SPM has been widely used for battery state estimation and prediction [18–20], charge/discharge optimization [21,22], and battery parameter estimation [23–26]. However, a major limitation of the SPM is that the underlying assumptions are only valid under low charge/discharge C-rates (e.g., less than 0.3 C). The typical operational C-rates in electrical vehicles and applications such as drones and various devices often exceed the capability of SPM. The gradient of lithium-ion concentration and electrolyte potential cause the SPM to give a large error.

Several approaches have been proposed to overcome the limitation of SPM while maintaining high computational efficiency. One method is to use transfer functions to analytically solve the electrolyte concentration distribution along the thickness direction of a battery cell [27–30]. For instance, Xie et al. revised the SPM by the cosine approximation method to calculate the electrolyte concentration [27]. They firstly applied Laplace transform to solve the diffusion equations analytically, and then obtained the complete transfer function. Another approach is to use some forms of polynomials to represent the electrolyte concentration and potential distribution within the cell [31–35]. For instance, Rahimian et al. developed an extended SPM by expressing the electrolyte concentration and potential in the electrode regions with cubic polynomials, and expressing the electrolyte concentration and potential in the separator region with parabolas [35]. Mehta et al. developed an extended SPM by applying second-order polynomials to express the electrolyte concentration, electrolyte potential, and solid phase potential distribution [34]. They formed their polynomials as a linear-space system for obtaining the solution. Li et al. developed a reduced-order electrochemical model by using polynomials to express the electrolyte concentration distribution in two divided calculation domains [33]. They combined the simplified electrolyte diffusion and other dynamics to form a five-state diagonal system. These studies show that polynomials can effectively represent the polarization distribution of electrolyte concentration and potential within a battery cell to improve the SPM for better accuracy while maintaining a high computational speed.

The demands for improving the present physics-based, reduced-order models include better accuracy at higher C-rates, capturing complex charging/discharging dynamics (e.g., dynamic driving profiles), and capability to systematically add new reactions such as aging mechanisms. These drive us to revisit the SPM to propose a different assumption for improved accuracy, and go beyond any SPM assumption by developing a reduced-order model directly from the P2D model. Polynomials have demonstrated great potentials in describing the electrolyte concentration and potential distribution, however, a major issue is that current approaches require subjectively selecting a few points (often selected without justification) along the electrode thickness to match various equations (e.g., reaction current density, diffusion) at those selected points, so as to construct the necessary number of relations for determining the unknown polynomial coefficients. Therefore, depending on where these points are selected, the results can be quite different even for the same problem. Thus, a fundamental self-consistent approach without requiring subjectively selecting "matching points" is needed.

We propose two physics-based, reduce-order models, namely revised single-particle model (RSPM) and fast-calculating P2D model (FCP2D). They are related, with RSPM being faster and suitable for moderately high C-rates, while FCP2D being more accurate at very high C-rates. For both models, we use higher-order polynomials to express the lithium-ion concentration and potential distribution. This capturing of non-uniform lithium-ion concentration overcomes the limitation of the traditional SPM. In the RSPM, the interfacial current density is assumed to be uniformly distributed at each moment within the cathode and the anode regions, thus the cathode and anode can be viewed as "single particles" from the perspective of interfacial current density. This treatment accelerates the calculation. In the FCP2D, this assumption is removed. Notably, we develop an innovative weak form of equations by using shape functions, which allow determining the evolution of polynomial coefficients self-consistently without relying on any matching points. This approach provides high robustness, especially for higher-order polynomials. We show that the proposed models can accurately calculate various electrochemical parameters (e.g., terminal voltage, lithium-ion concentration, electrolyte potential, lithium concentration on the particle surface, and interfacial current density) with high speeds, for both constant current charging/discharging and dynamic driving conditions.

## 2. Methodology

*2.1. Electrochemical model*

**Fig. 1** shows a schematic of the battery cell, which consists of an anode, a separator, a cathode, and two current collectors. The thickness of the anode (negative electrode), the separator, and the cathode (positive electrode) is denoted by $L_n$, $L_s$ and $L_p$, respectively. The materials of the anode and cathode particles are selected to be graphite and NMC 811, respectively. The electrolyte is selected to be $LiPF_6$ in EC:DEC (1:1 vol.%).

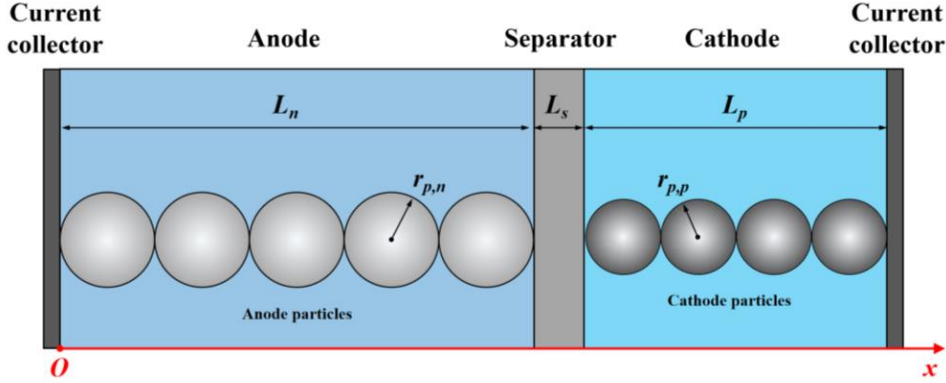

**Fig. 1.** Schematic of a lithium-ion battery cell.

The pseudo 2D (P2D) electrochemical model is widely used for battery simulations. In the P2D model, the potential of the solid phase and the electrolyte phase are governed by

$$\nabla \cdot \left( \sigma_{s,i}^{eff} \nabla \phi_{s,i} \right) = a_{s,i} i_{loc,i}, \tag{1}$$

$$\nabla \cdot \left( -\kappa_{e,i}^{eff} \left( \nabla \phi_{e,i} - \beta \nabla \ln c_{e,i} \right) \right) = a_{s,i} i_{loc,i}, \tag{2}$$

where $i$ denotes the region in a battery cell with $i = n$ for the negative electrode region, $i = s$ for the separator region, and $i = p$ for the positive electrode region. $\phi_{s,i}$ and $\phi_{e,i}$ denote the solid phase potential and the electrolyte phase potential, respectively. $\sigma_{s,i}^{eff}$ and $\kappa_{e,i}^{eff}$ denote the effective solid phase and electrolyte phase conductivity, respectively. They can be expressed by and $\kappa_{e,i}^{eff} = \kappa_{e0} \varepsilon_{e,i}^{burg}$, with $\sigma_{s0,i}$ and $\kappa_{e0}$ being the bulk solid phase and bulk electrolyte phase conductivity, $\varepsilon_{s,i}$ and $\varepsilon_{e,i}$ being the solid phase volume fraction and the electrolyt $\sigma_{s,i}^{eff} = \sigma_{s0,i} \varepsilon_{s,i}^{burg}$ e phase volume fraction, and *burg* being the Bruggeman constant. $a_{s,i}$ denotes the active surface area per electrode volume, which is given by $a_{s,i} = 3\varepsilon_{s,i} / r_{p,i}$ with $r_{p,i}$ being the particle radius. $i_{loc,i}$ denotes the interfacial current density. $c_{e,i}$ denotes the lithium-ion concentration in the electrolyte phase. The parameter $\beta$ is defined by

$$\beta = \frac{2RT}{F} \left( 1 + \frac{d \ln f_{\pm}}{d \ln c_{e,i}} \right) \left( 1 - t_{+}^{0} \right). \tag{3}$$

where $R$ denotes ideal gas constant, $T$ denotes temperature, $F$ denotes Faraday's constant, $f_{\pm}$ denotes the electrolyte activity coefficient for which we follow the common assumption of $d \ln f_{\pm} / d \ln c_{e,i} = 0$. $t_{+}^{0}$ denotes the lithium-ion transference number.

The lithium-ion concentration in the electrolyte is governed by

$$\frac{\partial \left( \varepsilon_{e,i} c_{e,i} \right)}{\partial t} = \nabla \cdot \left( D_{e,i}^{eff} \nabla c_{e,i} \right) + \frac{\left(1 - t_{+}^{0}\right)}{F} a_{s,i} i_{loc,i}, \tag{4}$$

where $t$ denotes time, and $D_{e,i}^{eff}$ denotes the effective electrolyte diffusion coefficient, which is given by $D_{e,i}^{eff} = D_{e0} \varepsilon_{e,i}^{burg}$ with $D_{e0}$ being the lithium-ion diffusion coefficient in bulk electrolyte.

The lithium concentration in each particle is governed by

$$\frac{\partial c_{s,i}}{\partial t} = \frac{1}{r^2} \frac{\partial}{\partial r} \left( D_{s,i} r^2 \frac{\partial c_{s,i}}{\partial r} \right), \tag{5}$$

where $c_{s,i}$ denotes the lithium concentration in the particles of electrodes, $r$ denotes the radial coordinate, and $D_{s,i}$ denotes the lithium diffusion coefficient in the particles of the solid phase.

The interfacial current density, $i_{loc,i}$, is given by the Bulter-Volmer equation,

$$i_{loc,i} = F k_i \left( c_{s,max,i} - c_{s,surf,i} \right)^{\alpha} c_{s,surf,i}^{1-\alpha} c_{e,i}^{\alpha} \left[ \exp\left( \frac{\alpha F \eta_i}{RT} \right) - \exp\left( -\frac{(1-\alpha) F \eta_i}{RT} \right) \right], \tag{6}$$

where $k_i$ denotes the reaction rate constant, $c_{s,max,i}$ denotes the maximum lithium ion concentration in the electrode particle, $c_{s,surf,i}$ denotes the surface lithium ion concentration, $\alpha$ denotes the anodic charge transfer coefficient, and $\eta_i$ denotes the over-potential. The over-potential is given by

$$\eta_i = \phi_{s,i} - \phi_{e,i} - U_{eq,i}, \tag{7}$$

where $U_{eq,i}$ denotes the equilibrium potential.

The boundary conditions for the particles are given by

$$\left. \frac{\partial c_{s,i}}{\partial r} \right|_{r=0} = 0, \; \left. D_{s,i} \frac{\partial c_{s,i}}{\partial r} \right|_{r=r_{p,i}} = -\frac{i_{loc,i}}{F}. \tag{8}$$

The boundary conditions for the lithium-ion concentration in the electrode level are given by

$$\left. \frac{\partial c_{e,n}}{\partial x} \right|_{x=0} = 0, \; \left. \frac{\partial c_{e,p}}{\partial x} \right|_{x=L_n+L_s+L_p} = 0. \tag{9}$$

The boundary conditions for the electrolyte potential in the electrode level are given by

$$\left.\frac{\partial \phi_{e,n}}{\partial x}\right|_{x=0} = 0, \quad \left.\frac{\partial \phi_{e,p}}{\partial x}\right|_{x=L_n+L_s+L_p} = 0 \quad . \tag{10}$$

The boundary conditions for the solid potential in the electrode level are given by

$$\left.\sigma_{s,p}^{eff} \frac{\partial \phi_{s,p}}{\partial x}\right|_{x=L_n+L_s+L_p} = -i_{app} \quad , \tag{11}$$

where $i_{app}$ denotes the applied current density ($i_{app} > 0$ for discharge). In principle, one can choose any point in the battery to define the zero potential. As shown later, for our approach it is more convenient to choose the electrolyte potential at $x = 0$ to define the zero potential.

The continuity conditions for the lithium-ion concentration and the electrolyte potential are given by

$$\left.c_{e,n}\right|_{x=L_n} = \left.c_{e,s}\right|_{x=L_n}, \quad \left.c_{e,s}\right|_{x=L_n+L_s} = \left.c_{e,p}\right|_{x=L_n+L_s}, \tag{12}$$

$$\left.\phi_{e,n}\right|_{x=L_n} = \left.\phi_{e,s}\right|_{x=L_n}, \quad \left.\phi_{e,s}\right|_{x=L_n+L_s} = \left.\phi_{e,p}\right|_{x=L_n+L_s}. \tag{13}$$

The continuity conditions for the electrolyte current density, $\mathbf{i}_{e,i} = -\kappa_{e,i}^{eff}\left(\nabla \phi_{e,i} - \beta \nabla \ln c_{e,i}\right)$, are given by

$$\left.-\kappa_{e,n}^{eff}\left(\frac{\partial \phi_{e,n}}{\partial x} - \beta \frac{\partial \ln c_{e,n}}{\partial x}\right)\right|_{x=L_n} = \left.-\kappa_{e,s}^{eff}\left(\frac{\partial \phi_{e,s}}{\partial x} - \beta \frac{\partial \ln c_{e,s}}{\partial x}\right)\right|_{x=L_n}, \tag{14}$$

$$\left.-\kappa_{e,s}^{eff}\left(\frac{\partial \phi_{e,s}}{\partial x} - \beta \frac{\partial \ln c_{e,s}}{\partial x}\right)\right|_{x=L_n+L_s} = \left.-\kappa_{e,p}^{eff}\left(\frac{\partial \phi_{e,p}}{\partial x} - \beta \frac{\partial \ln c_{e,p}}{\partial x}\right)\right|_{x=L_n+L_s}. \tag{15}$$

The continuity condition for the lithium ion flux, $\mathbf{N}_{e,i} = -D_{e,i}^{eff}\nabla c_{e,i} + \mathbf{i}_{e,i} t_+^0 / F$, is given by

$$\left.-D_{e,n}^{eff}\frac{\partial c_{e,n}}{\partial x}\right|_{x=L_n} = \left.-D_{e,s}^{eff}\frac{\partial c_{e,s}}{\partial x}\right|_{x=L_n}, \tag{16}$$

$$\left.-D_{e,s}^{eff}\frac{\partial c_{e,s}}{\partial x}\right|_{x=L_n+L_s} = \left.-D_{e,p}^{eff}\frac{\partial c_{e,p}}{\partial x}\right|_{x=L_n+L_s}. \tag{17}$$

Note that in the above equations the term $\mathbf{i}_{e,i} t_+^0 / F$ does not show up since we have used the continuity condition for $\mathbf{i}_{e,i}$.

## 2.2. Reduced-order model for fast calculation

In this section, we develop two reduce-order physical-based models based on the full electrochemical equations in Section 2.1. In the RSPM, the interfacial current density is assumed to be uniformly distributed at each moment within the cathode and the anode regions, thus the cathode and anode can be viewed as single particles. This treatment accelerates the calculation. In the FCP2D, this assumption on interfacial current density uniformity is removed, thus the cathode and anode cannot be modeled as single particles and the pseudo dimension is needed (which is the same as the P2D electrochemical model, as illustrated in **Fig. 1**).

For both the RSPM and the FCP2D, the lithium-ion concentration and the electrolyte potential distribution along the thickness direction of the battery cell are assumed to be polynomial functions,

$$c_{e,i} = a_{i,0}(t) + a_{i,1}(t)x_i + a_{i,2}(t)x_i^2 + a_{i,3}(t)x_i^3, \tag{18}$$

$$\phi_{e,i} = b_{i,0}(t) + b_{i,1}(t)x_i + b_{i,2}(t)x_i^2 + b_{i,3}(t)x_i^3, \tag{19}$$

where $a_i(t)$ and $b_i(t)$ denote the time-dependent polynomial parameters, and $x_i$ denotes the normalized position in region $i$ along the thickness direction of the battery cell calculated by

$$x_n = \frac{x}{L_n}, \quad x_s = \frac{x - L_n}{L_s}, \quad x_p = \frac{x - L_n - L_s}{L_p}, \tag{20}$$

Based on Eq. (18), the specific expressions for the lithium ion concentration in the electrolyte in the anode, separator, and cathode regions are

$$\begin{aligned} c_{e,n} &= a_{n,0} + a_{n,1}x_n + a_{n,2}x_n^2 + a_{n,3}x_n^3 \\ c_{e,s} &= a_{s,0} + a_{s,1}x_s + a_{s,2}x_s^2 \\ c_{e,p} &= a_{p,0} + a_{p,1}x_p + a_{p,2}x_p^2 + a_{p,3}x_p^3 \end{aligned}. \tag{21}$$

With Eq. (20), the boundary conditions in Eq. (9) become

$$\left.\frac{\partial c_{e,n}}{\partial x_n}\right|_{x_n=0} = 0, \quad \left.\frac{\partial c_{e,p}}{\partial x_p}\right|_{x_p=1} = 0. \tag{22}$$

The continuity conditions in Eqs. (12), (16), (17) become

$$c_{e,n}\big|_{x_n=1} = c_{e,s}\big|_{x_s=0}, \quad c_{e,p}\big|_{x_p=0} = c_{e,s}\big|_{x_s=1}, \tag{23}$$

$$-D_{e,n}^{eff}\frac{\partial c_{e,n}}{L_n \partial x_n}\bigg|_{x_n=1} = -D_{e,s}^{eff}\frac{\partial c_{e,s}}{L_s \partial x_s}\bigg|_{x_s=0}, \tag{24}$$

$$-D_{e,p}^{eff} \frac{\partial c_{e,p}}{L_p \partial x_p}\bigg|_{x_p=0} = -D_{e,s}^{eff} \frac{\partial c_{e,s}}{L_s \partial x_s}\bigg|_{x_s=1}. \tag{25}$$

Substituting Eq. (21) into the 6 constrains in Eqs. (22)−(25), we can reduce the number of unknown polynomial parameters from 11 to 5. Choosing $a_{n,0}$, $a_{n,2}$, $a_{n,3}$, $a_{s,2}$ and $a_{p,3}$ as the independent variables, we get

$$a_{n,1} = 0, \tag{26}$$

$$a_{s,0} = a_{n,0} + a_{n,2} + a_{n,3}, \tag{27}$$

$$a_{s,1} = 2R_1 a_{n,2} + 3R_1 a_{n,3}, \tag{28}$$

$$a_{p,0} = a_{n,0} + (1+2R_1)a_{n,2} + (1+3R_1)a_{n,3} + a_{s,2}, \tag{29}$$

$$a_{p,1} = 2R_1 R_2 a_{n,2} + 3R_1 R_2 a_{n,3} + 2R_2 a_{s,2}, \tag{30}$$

$$a_{p,2} = -R_1 R_2 a_{n,2} - \frac{3}{2} R_1 R_2 a_{n,3} - R_2 a_{s,2} - \frac{3}{2} a_{p,3}, \tag{31}$$

where $R_1$ and $R_2$ are defined as

$$R_1 = \frac{L_s}{L_n}\left(\frac{\varepsilon_{e,n}}{\varepsilon_{e,s}}\right)^{burg}, \quad R_2 = \frac{L_p}{L_s}\left(\frac{\varepsilon_{e,s}}{\varepsilon_{e,p}}\right)^{burg}. \tag{32}$$

Based on Eq. (19), the specific expressions for the electrolyte potential in the anode and cathode regions are

$$\begin{aligned}\phi_{e,n} &= b_{n,0} + b_{n,1}x_n + b_{n,2}x_n^2 + b_{n,3}x_n^3 \\ \phi_{e,p} &= b_{p,0} + b_{p,1}x_p + b_{p,2}x_p^2 + b_{p,3}x_p^3\end{aligned}. \tag{33}$$

With Eq. (20), the boundary and continuity conditions in Eqs. (10), (13)−(15) become

$$\frac{\partial \phi_{e,n}}{\partial x_n}\bigg|_{x_n=0} = 0, \quad \frac{\partial \phi_{e,p}}{\partial x_p}\bigg|_{x_p=1} = 0, \tag{34}$$

$$\phi_{e,n}\big|_{x_n=1} = \phi_{e,s}\big|_{x_s=0}, \quad \phi_{e,s}\big|_{x_s=1} = \phi_{e,p}\big|_{x_p=0}, \tag{35}$$

$$-\frac{\kappa_{e,n}^{eff}}{L_n}\left(\frac{\partial \phi_{e,n}}{\partial x_n} - \beta \frac{\partial \ln c_{e,n}}{\partial x_n}\right)\bigg|_{x_n=1} = -\frac{\kappa_{e,s}^{eff}}{L_s}\left(\frac{\partial \phi_{e,s}}{\partial x_s} - \beta \frac{\partial \ln c_{e,s}}{\partial x_s}\right)\bigg|_{x_s=0} = i_{app}, \tag{36}$$

$$-\frac{\kappa_{e,s}^{eff}}{L_s}\left(\frac{\partial \phi_{e,s}}{\partial x_s} - \beta \frac{\partial \ln c_{e,s}}{\partial x_s}\right)\bigg|_{x_s=1} = -\frac{\kappa_{e,p}^{eff}}{L_p}\left(\frac{\partial \phi_{e,p}}{\partial x_p} - \beta \frac{\partial \ln c_{e,p}}{\partial x_p}\right)\bigg|_{x_p=0} = i_{app}. \tag{37}$$

In Eqs. (36) and (37) we have included the relation between the electrolyte current density and the applied current density, $i_{app}$. The conditions in Eq. (34)−(37) can be used to reduce the number of unknown polynomial parameters and to solve the electrolyte potential in the separator region.

The electrolyte potential is defined relative to a reference. For convenience, we choose the 0 potential by

$$\phi_{e,n}\big|_{x_n=0} = 0. \tag{38}$$

Eqs. (36) and (37) give

$$-\left(\frac{\partial \phi_{e,n}}{\partial x_n} - \beta \frac{\partial \ln c_{e,n}}{\partial x_n}\right)\bigg|_{x_n=1} = \frac{L_n i_{app}}{\kappa_{e,n}^{eff}}, \tag{39}$$

$$-\left(\frac{\partial \phi_{e,p}}{\partial x_p} - \beta \frac{\partial \ln c_{e,p}}{\partial x_p}\right)\bigg|_{x_p=0} = \frac{L_p i_{app}}{\kappa_{e,p}^{eff}}, \tag{40}$$

Substituting Eq. (33) into the 5 constrains in Eqs. (34), (38)−(40), we can reduce the number of unknown polynomial parameters from 8 to 3. Choosing $b_{n,3}$, $b_{p,0}$ and $b_{p,3}$ as the independent variables, we get

$$b_{n,0} = 0, \tag{41}$$

$$b_{n,1} = 0, \tag{42}$$

$$b_{n,2} = \frac{\beta}{2} \frac{2a_{n,2} + 3a_{n,3}}{a_{n,0} + a_{n,2} + a_{n,3}} - \frac{L_n i_{app}}{2\kappa_{e,n}^{eff}} - \frac{3}{2}b_{n,3}, \tag{43}$$

$$b_{p,1} = \beta \frac{a_{p,1}}{a_{p,0}} - \frac{L_p i_{app}}{\kappa_{e,p}^{eff}}, \tag{44}$$

$$b_{p,2} = -\frac{\beta}{2} \frac{a_{p,1}}{a_{p,0}} + \frac{L_p i_{app}}{2\kappa_{e,p}^{eff}} - \frac{3}{2}b_{p,3}. \tag{45}$$

The electrolyte potential in the separator region is solved by letting the right-hand side of Eq. (2) to be 0, i.e.,

$$\frac{\partial}{\partial x_s}\left(\frac{\partial \phi_{e,s}}{\partial x_s} - \beta \frac{\partial \ln c_{e,s}}{\partial x_s}\right) = 0. \tag{46}$$

Integrating Eq. (46) and applying the conditions of Eqs. (35)−(37) (see detailed derivation in **Appendix A**), we get

$$\phi_{e,s} = \beta \ln\left(a_{s,0} + a_{s,1}x_s + a_{s,2}x_s^2\right) - \frac{L_s i_{app}}{\kappa_{e,s}^{eff}} x_s + \left(b_{n,2} + b_{n,3} - \beta \ln\left(a_{s,0}\right)\right), \tag{47}$$

and the relation

$$b_{p,0} = \beta \ln\left(\frac{a_{s,0} + a_{s,1} + a_{s,2}}{a_{s,0}}\right) - \left(\frac{L_s}{\kappa_{e,s}^{eff}} + \frac{L_n}{2\kappa_{e,n}^{eff}}\right)i_{app} + \frac{\beta}{2}\frac{2a_{n,2} + 3a_{n,3}}{a_{n,0} + a_{n,2} + a_{n,3}} - \frac{b_{n,3}}{2}, \tag{48}$$

Therefore, we have only two independent variables, $b_{n,3}$ and $b_{p,3}$. Eqs. (41)−(45) and (48) give the other polynomial parameters.

We need to solve the 5 unknown polynomial parameters ($a_{n,0}$, $a_{n,2}$, $a_{n,3}$, $a_{p,3}$, $a_{s,2}$) for the lithium ion concentration and 2 unknown polynomial parameters ($b_{n,3}$, $b_{p,3}$) for the electrolyte potential. Here we proposal a new approach of using shape functions (kernel functions) to determine the unknown polynomial parameters, and key idea is to construct a weak integration form.

To introduce the approach, consider solving an equation

$$f(x) = 0. \tag{49}$$

An equivalent weak form (integration form) is constructed by using a shape function (kernel function), $w(x)$,

$$\int w(x) \cdot f(x) = 0 \tag{50}$$

The shape function is given by

$$w(x) = d_0 + w_1 x_i + w_2 x_i^2 + w_3 x_i^3, \tag{51}$$

where $d_0$ denotes the bias while $w_1$, $w_2$, $w_3$ denote three weights. Their values are pre-selected and known. In principle, any selected values are fine. One can also use optimization to determine a better choice of the bias and weight values.

Now we construct the weak form for Eqs. (2) and (4), which govern the electrolyte potential and concentration. We get

$$\int_0^1 w(x)\frac{\partial c_{e,i}}{\partial t}dx_i = \frac{D_{e,i}^{eff}}{\varepsilon_{e,i}L_i^2}\int_0^1 w(x)\frac{\partial^2 c_{e,i}}{\partial x_i^2}dx_i + \frac{(1-t_+^0)a_{s,i}}{\varepsilon_{e,i}F}\int_0^1 w(x)i_{loc,i}dx_i, \tag{52}$$

$$-\int_0^1 w(x)\frac{\partial^2 \phi_{e,i}}{\partial x_i^2}dx_i + \beta\int_0^1 w(x)\frac{\partial}{\partial x_i}\left(\frac{1}{c_{e,i}}\frac{\partial c_{e,i}}{\partial x_i}\right)dx_i = \frac{a_{s,i}L_i^2}{\kappa_{e,i}^{eff}}\int_0^1 w(x)i_{loc,i}dx_i, \qquad (53)$$

We firstly set the bias $d_0 = 1$ and $w_1 = w_2 = w_3 = 0$ to construct 3 equations for solving the unknown polynomial parameters in lithium ion concentration, which give

$$\int_0^1\left(\frac{\partial c_{e,n}}{\partial t}\right)dx_n = \frac{D_{e,n}^{eff}}{\varepsilon_{e,n}L_n^2}\int_0^1\left(\frac{\partial^2 c_{e,n}}{\partial x_n^2}\right)dx_n + \frac{(1-t_+^0)a_{s,n}}{\varepsilon_{e,n}F}\int_0^1 i_{loc,n}dx_n, \qquad (54)$$

$$\int_0^1\left(\frac{\partial c_{e,s}}{\partial t}\right)dx_s = \frac{D_{e,s}^{eff}}{\varepsilon_{e,s}L_s^2}\int_0^1\left(\frac{\partial^2 c_{e,s}}{\partial x_s^2}\right)dx_s, \qquad (55)$$

$$\int_0^1\left(\frac{\partial c_{e,p}}{\partial t}\right)dx_p = \frac{D_{e,p}^{eff}}{\varepsilon_{e,p}L_p^2}\int_0^1\left(\frac{\partial^2 c_{e,p}}{\partial x_p^2}\right)dx_p + \frac{(1-t_+^0)a_{s,p}}{\varepsilon_{e,p}F}\int_0^1 i_{loc,p}dx_p. \qquad (56)$$

The total reaction current in the anode and cathod relates to $i_{app}$ by

$$\int_0^1 a_{s,n}i_{loc,n}dx_n = \frac{i_{app}}{L_n}, \quad \int_0^1 a_{s,p}i_{loc,p}dx_p = -\frac{i_{app}}{L_p}. \qquad (57)$$

Substituting the electrolyte concentration in Eq. (21), the relations in Eq. (26)−(31), and Eq. (57) into Eqs. (54)−(56), we get

$$\frac{da_{n,0}}{dt} + \frac{1}{3}\frac{da_{n,2}}{dt} + \frac{1}{4}\frac{da_{n,3}}{dt} = P_n(2a_{n,2} + 3a_{n,3}) + Q_n i_{app}, \qquad (58)$$

$$\frac{da_{n,0}}{dt} + (1+R_1)\frac{da_{n,2}}{dt} + \left(1+\frac{3}{2}R_1\right)\frac{da_{n,3}}{dt} + \frac{1}{3}\frac{da_{s,2}}{dt} = 2P_s a_{s,2}, \qquad (59)$$

$$\frac{da_{n,0}}{dt} + \left(1+2R_1 + \frac{2}{3}R_1R_2\right)\frac{da_{n,2}}{dt} + (1+3R_1 + R_1R_2)\frac{da_{n,3}}{dt} + \left(1+\frac{2}{3}R_2\right)\frac{da_{s,2}}{dt} - \frac{1}{4}\frac{da_{p,3}}{dt}$$
$$= -P_p(2R_1R_2 a_{n,2} + 3R_1R_2 a_{n,3} + 2R_2 a_{s,2}) - Q_p I_{app} \qquad (60)$$

where

$$P_n = \frac{D_{e,n}^{eff}}{\varepsilon_{e,n}L_n^2}, \quad P_s = \frac{D_{e,s}^{eff}}{\varepsilon_{e,s}L_s^2}, \quad P_p = \frac{D_{e,p}^{eff}}{\varepsilon_{e,p}L_p^2}, \quad Q_n = \frac{(1-t_+^0)}{\varepsilon_{e,n}FL_n}, \quad Q_p = \frac{(1-t_+^0)}{\varepsilon_{e,p}FL_p}. \qquad (61)$$

Eqs. (58)−(60) give 3 ordinary differential equations involving the 5 independent variables $a_{n,0}$, $a_{n,2}$, $a_{n,3}$, $a_{s,2}$ and $a_{p,3}$.

The weak form of Eq. (53) with $d_0 = 1$ and $w_1 = w_2 = w_3 = 0$ gives the same equations as Eqs. (39) and (40), which have already been used.

Next, we set the bias $d_0 = 0$ and use $w_1$, $w_2$, $w_3$ (at least one of them is non-zero) to construct 4 more equations. The expression of the local interfacial current density in RSPM and FCP2D are different, leading to different ways to construct the remaining equations. So in the following we separate them into two subsections.

2.2.1. *Revised single-particle model (RSPM)*

For RSPM, we assume the interfacial current density to be uniformly distributed along the thickness direction of the cathode or the anode. This assumption allows obtaining an analytical form of the interfacial current density,

$$i_{loc,n,RSPM} = \frac{i_{app}}{a_{s,n} L_n}, \quad i_{loc,p,RSPM} = -\frac{i_{app}}{a_{s,p} L_p}. \tag{62}$$

Substituting the electrolyte concentration in Eq. (21), the relations in Eq. (26)−(31), $w(x) = w_1 x_i + w_2 x_i^2 + w_3 x_i^3$ and Eq. (62) into Eqs. (52), we get

$$\left(\frac{1}{2}w_1 + \frac{1}{3}w_2 + \frac{1}{4}w_3\right)\frac{da_{n,0}}{dt} + \left(\frac{1}{4}w_1 + \frac{1}{5}w_2 + \frac{1}{6}w_3\right)\frac{da_{n,2}}{dt} + \left(\frac{1}{5}w_1 + \frac{1}{6}w_2 + \frac{1}{7}w_3\right)\frac{da_{n,3}}{dt}$$
$$= w_1 P_n \left(a_{n,2} + 2a_{n,3}\right) + w_2 P_n \left(\frac{2}{3}a_{n,2} + \frac{3}{2}a_{n,3}\right) + w_3 P_n \left(\frac{1}{2}a_{n,2} + \frac{6}{5}a_{n,3}\right) \tag{63}$$
$$+ \left(\frac{1}{2}w_1 + \frac{1}{3}w_2 + \frac{1}{4}w_3\right) Q_n i_{app}$$

and

$$\left(\frac{1}{2}w_1 + \frac{1}{3}w_2 + \frac{1}{4}w_3\right)\frac{da_{n,0}}{dt}$$
$$+\left(\left(\frac{1}{2}+R_1+\frac{5}{12}R_1R_2\right)w_1+\left(\frac{1}{3}+\frac{2R_1}{3}+\frac{3}{10}R_1R_2\right)w_2+\left(\frac{1}{4}+\frac{R_1}{2}+\frac{7}{30}R_1R_2\right)w_3\right)\frac{da_{n,2}}{dt}$$
$$+\left(\left(\frac{1}{2}+\frac{3R_1}{2}+\frac{5}{8}R_1R_2\right)w_1+\left(\frac{1}{3}+R_1+\frac{9}{20}R_1R_2\right)w_2+\left(\frac{1}{4}+\frac{3R_1}{4}+\frac{7}{20}R_1R_2\right)w_3\right)\frac{da_{n,3}}{dt}$$
$$+\left(\left(\frac{1}{2}+\frac{5}{12}R_2\right)w_1+\left(\frac{1}{3}+\frac{3}{10}R_2\right)w_2+\left(\frac{1}{4}+\frac{7}{30}R_2\right)w_3\right)\frac{da_{s,2}}{dt} \quad (64)$$
$$-\left(\frac{7}{40}w_1+\frac{2}{15}w_2+\frac{3}{28}w_3\right)\frac{da_{p,3}}{dt}$$
$$=-\left(w_1+\frac{2}{3}w_2+\frac{1}{2}w_3\right)R_1R_2P_pa_{n,2}-\left(\frac{3}{2}w_1+w_2+\frac{3}{4}w_3\right)R_1R_2P_pa_{n,3}-\left(w_1+\frac{2}{3}w_2+\frac{1}{2}w_3\right)R_2P_pa_{s,2}$$
$$+\left(\frac{1}{2}w_1+\frac{1}{2}w_2+\frac{9}{20}w_3\right)P_pa_{p,3}-\left(\frac{1}{2}w_1+\frac{1}{3}w_2+\frac{1}{4}w_3\right)Q_pi_{app}$$

The 5 ordinary differential equations Eqs. (58)−(60), (63), (64) can be used to solve $a_{n,0}$, $a_{n,2}$, $a_{n,3}$, $a_{s,2}$ and $a_{p,3}$. The numerical calculation is fast since we only need to solve a group of ordinary differential equations.

Next, we construct the 2 equations to solve $b_{n,3}$, $b_{p,3}$ for the electrolyte potential. Substituting the electrolyte potential in Eq. (33), $w(x) = w_1x_i + w_2x_i^2 + w_3x_i^3$ and Eq. (62) into Eq. (53), we get

$$\beta\int_0^1 \left(w_1x_n+w_2x_n^2+w_3x_n^3\right)\left(\frac{(2a_{n,2}+6a_{n,3}x_n)(a_{n,0}+a_{n,1}x_n+a_{n,2}x_n^2+a_{n,3}x_n^3)-(a_{n,1}+2a_{n,2}x_n+3a_{n,3}x_n^2)^2}{(a_{n,0}+a_{n,1}x_n+a_{n,2}x_n^2+a_{n,3}x_n^3)^2}\right)dx_n,$$
$$= w_1(b_{n,2}+2b_{n,3})+w_2\left(\frac{2}{3}b_{n,2}+\frac{3}{2}b_{n,3}\right)+w_3\left(\frac{1}{2}b_{n,2}+\frac{6}{5}b_{n,3}\right)+\left(\frac{1}{2}w_1+\frac{1}{3}w_2+\frac{1}{4}w_3\right)\frac{L_ni_{app}}{\kappa_{e,n}^{eff}} \quad (65)$$

$$\beta\int_0^1 \left(w_1x_p+w_2x_p^2+w_3x_p^3\right)\left(\frac{(2a_{p,2}+6a_{p,3}x_p)(a_{p,0}+a_{p,1}x_p+a_{p,2}x_p^2+a_{p,3}x_p^3)-(a_{p,1}+2a_{p,2}x_p+3a_{p,3}x_p^2)^2}{(a_{p,0}+a_{p,1}x_p+a_{p,2}x_p^2+a_{p,3}x_p^3)^2}\right)dx_p.$$
$$= w_1(b_{p,2}+2b_{p,3})+w_2\left(\frac{2}{3}b_{p,2}+\frac{3}{2}b_{p,3}\right)+w_3\left(\frac{1}{2}b_{p,2}+\frac{6}{5}b_{p,3}\right)-\left(\frac{1}{2}w_1+\frac{1}{3}w_2+\frac{1}{4}w_3\right)\frac{L_pi_{app}}{\kappa_{e,p}^{eff}} \quad (66)$$

Substituting Eqs. (43) and (45) into the above two equations, we can obtain the analytical expression for $b_{n,3}$, $b_{p,3}$

$$b_{n,3} = \frac{20\beta}{10w_1 + 10w_2 + 9w_3} \times \left[ \int_0^1 (w_1 x_n + w_2 x_n^2 + w_3 x_n^3) \left( \frac{(2a_{n,2} + 6a_{n,3} x_n)(a_{n,0} + a_{n,1} x_n + a_{n,2} x_n^2 + a_{n,3} x_n^3) - (a_{n,1} + 2a_{n,2} x_n + 3a_{n,3} x_n^2)^2}{(a_{n,0} + a_{n,1} x_n + a_{n,2} x_n^2 + a_{n,3} x_n^3)^2} \right) dx_n \right.$$

$$\left. - \left( \frac{1}{2} w_1 + \frac{1}{3} w_2 + \frac{1}{4} w_3 \right) \frac{2a_{n,2} + 3a_{n,3}}{a_{n,0} + a_{n,2} + a_{n,3}} \right]$$

(67)

$$b_{p,3} = \frac{20\beta}{10w_1 + 10w_2 + 9w_3} \times \left[ \int_0^1 (w_1 x_p + w_2 x_p^2 + w_3 x_p^3) \left( \frac{(2a_{p,2} + 6a_{p,3} x_p)(a_{p,0} + a_{p,1} x_p + a_{p,2} x_p^2 + a_{p,3} x_p^3) - (a_{p,1} + 2a_{p,2} x_p + 3a_{p,3} x_p^2)^2}{(a_{p,0} + a_{p,1} x_p + a_{p,2} x_p^2 + a_{p,3} x_p^3)^2} \right) dx_p \right.$$

$$\left. + \left( \frac{1}{2} w_1 + \frac{1}{3} w_2 + \frac{1}{4} w_3 \right) \frac{a_{p,1}}{a_{p,0}} \right]$$

(68)

For analytically solving the solid potential of the anode and cathode in RSPM, we use the Bulter-Volmer equation

$$i_{loc,i,RSPM} = i_{0,i,RSPM} \left[ \exp\left( \frac{\alpha F \eta_{i,RSPM}}{RT} \right) - \exp\left( -\frac{(1-\alpha) F \eta_{i,RSPM}}{RT} \right) \right], \quad (69)$$

where

$$i_{0,i,RSPM} = F k_i \left( c_{s,\max,i} - c_{s,surf,i} \right)^{\alpha} \left( c_{s,surf,i} \right)^{1-\alpha} \left( \overline{c}_{e,i,ave} \right)^{\alpha}, \quad (70)$$

$$\eta_{i,RSPM} = \phi_{s,i} - \overline{\phi}_{e,i,ave} - U_{eq,i}. \quad (71)$$

In the above expressions $\overline{c}_{e,i,ave}$ denotes the average lithium-ion concentration along the thickness direction of the anode or cathode given by

$$\overline{c}_{e,n,ave} = \frac{1}{L_n} \int_0^1 (a_{n,0} + a_{n,1} x_n + a_{n,2} x_n^2 + a_{n,3} x_n^3) dx_n = \frac{1}{L_n} \left( a_{n,0} + \frac{1}{2} a_{n,1} + \frac{1}{3} a_{n,2} + \frac{1}{4} a_{n,3} \right)$$

$$\overline{c}_{e,p,ave} = \frac{1}{L_p} \int_0^1 (a_{p,0} + a_{p,1} x_p + a_{p,2} x_p^2 + a_{p,3} x_p^3) dx_p = \frac{1}{L_p} \left( a_{p,0} + \frac{1}{2} a_{p,1} + \frac{1}{3} a_{p,2} + \frac{1}{4} a_{p,3} \right)$$

(72)

while $\overline{\phi}_{e,i,ave}$ denotes the average electrolyte potential along the thickness direction of the anode or cathode given by

$$\overline{\phi}_{e,n,ave} = \frac{1}{L_n} \int_0^1 (b_{n,0} + b_{n,1} x_n + b_{n,2} x_n^2 + b_{n,3} x_n^3) dx_n = \frac{1}{L_n} \left( b_{n,0} + \frac{1}{2} b_{n,1} + \frac{1}{3} b_{n,2} + \frac{1}{4} b_{n,3} \right)$$

$$\overline{\phi}_{e,p,ave} = \frac{1}{L_p} \int_0^1 (b_{p,0} + b_{p,1} x_p + b_{p,2} x_p^2 + b_{p,3} x_p^3) dx_p = \frac{1}{L_p} \left( b_{p,0} + \frac{1}{2} b_{p,1} + \frac{1}{3} b_{p,2} + \frac{1}{4} b_{p,3} \right)$$

(73)

We further define

$$m_n = i_{app} / F k_n a_{s,n} L_n \left(c_{s,\max,n} - c_{s,surf,n}\right)^\alpha \left(c_{s,surf,n}\right)^{1-\alpha} \left(\bar{c}_{e,n,ave}\right)^\alpha, \tag{74}$$

$$m_p = -i_{app} / F k_p a_{s,p} L_p \left(c_{s,\max,p} - c_{s,surf,p}\right)^\alpha \left(c_{s,surf,p}\right)^{1-\alpha} \left(\bar{c}_{e,p,ave}\right)^\alpha. \tag{75}$$

Substituting Eq. (62) into Eq. (69) gives $m_n = \exp\left(\alpha F \eta_{n,RSPM} / RT\right) - \exp\left(-(1-\alpha) F \eta_{n,RSPM} / RT\right)$ and $-m_p = \exp\left(\alpha F \eta_{p,RSPM} / (RT)\right) - \exp\left(-(1-\alpha) F \eta_{p,RSPM} / (RT)\right)$

Assuming $\alpha = 0.5$, we can solve the solid potential of the anode and cathode by

$$\phi_{s,n} = \eta_{n,RSPM} + \bar{\phi}_{e,n,ave} + U_{eq,n} = \frac{2RT}{F} \ln\left(\frac{\sqrt{m_n^2 + 4} + m_n}{2}\right) + \bar{\phi}_{e,n,ave} + U_{eq,n}, \tag{76}$$

$$\phi_{s,p} = \eta_{p,RSPM} + \bar{\phi}_{e,p,ave} + U_{eq,p} = -\frac{2RT}{F} \ln\left(\frac{\sqrt{m_p^2 + 4} + m_p}{2}\right) + \bar{\phi}_{e,p,ave} + U_{eq,p}, \tag{77}$$

The terminal voltage of the battery cell based on RSPM can then be calculated as

$$V_{t,RSPM} = U_{eq,p} - U_{eq,n} - \frac{2RT}{F} \ln\left(\frac{\sqrt{m_p^2 + 4} + m_p}{2}\right) - \frac{2RT}{F} \ln\left(\frac{\sqrt{m_n^2 + 4} + m_n}{2}\right) + \bar{\phi}_{e,p,ave} - \bar{\phi}_{e,n,ave}. \tag{78}$$

2.2.2. *Fast calculation P2D model (FCP2D)*

For RSPM, we assume the interfacial current density to be uniformly distributed along the thickness direction of the cathode or the anode. Here we remove this approximation to account for position-dependent interfacial current density.

Substituting the electrolyte concentration in Eq. (21), the relations in Eq. (26)−(31) and $w(x) = w_1 x_i + w_2 x_i^2 + w_3 x_i^3$ into Eqs. (52), we get

$$\left(\frac{1}{2}w_1 + \frac{1}{3}w_2 + \frac{1}{4}w_3\right)\frac{da_{n,0}}{dt} + \left(\frac{1}{4}w_1 + \frac{1}{5}w_2 + \frac{1}{6}w_3\right)\frac{da_{n,2}}{dt} + \left(\frac{1}{5}w_1 + \frac{1}{6}w_2 + \frac{1}{7}w_3\right)\frac{da_{n,3}}{dt}$$
$$= w_1 P_n \left(a_{n,2} + 2a_{n,3}\right) + w_2 P_n \left(\frac{2}{3}a_{n,2} + \frac{3}{2}a_{n,3}\right) + w_3 P_n \left(\frac{1}{2}a_{n,2} + \frac{6}{5}a_{n,3}\right) \tag{79}$$
$$+ Q_n L_n a_{s,n} \int_0^1 \left(w_1 x_n + w_2 x_n^2 + w_3 x_n^3\right) i_{loc,n,FCP2D} dx_n$$

and

$$\left(\frac{1}{2}w_1 + \frac{1}{3}w_2 + \frac{1}{4}w_3\right)\frac{da_{n,0}}{dt}$$

$$+\left(\left(\frac{1}{2}+R_1+\frac{5}{12}R_1R_2\right)w_1+\left(\frac{1}{3}+\frac{2R_1}{3}+\frac{3}{10}R_1R_2\right)w_2+\left(\frac{1}{4}+\frac{R_1}{2}+\frac{7}{30}R_1R_2\right)w_3\right)\frac{da_{n,2}}{dt}$$

$$+\left(\left(\frac{1}{2}+\frac{3R_1}{2}+\frac{5}{8}R_1R_2\right)w_1+\left(\frac{1}{3}+R_1+\frac{9}{20}R_1R_2\right)w_2+\left(\frac{1}{4}+\frac{3R_1}{4}+\frac{7}{20}R_1R_2\right)w_3\right)\frac{da_{n,3}}{dt}$$

$$+\left(\left(\frac{1}{2}+\frac{5}{12}R_2\right)w_1+\left(\frac{1}{3}+\frac{3}{10}R_2\right)w_2+\left(\frac{1}{4}+\frac{7}{30}R_2\right)w_3\right)\frac{da_{s,2}}{dt} \quad , \quad (80)$$

$$-\left(\frac{7}{40}w_1+\frac{2}{15}w_2+\frac{3}{28}w_3\right)\frac{da_{p,3}}{dt}$$

$$=-\left(w_1+\frac{2}{3}w_2+\frac{1}{2}w_3\right)R_1R_2P_p a_{n,2}-\left(\frac{3}{2}w_1+w_2+\frac{3}{4}w_3\right)R_1R_2P_p a_{n,3}-\left(w_1+\frac{2}{3}w_2+\frac{1}{2}w_3\right)R_2P_p a_{s,2}$$

$$+\left(\frac{1}{2}w_1+\frac{1}{2}w_2+\frac{9}{20}w_3\right)P_p a_{p,3}+Q_p L_p a_{s,p}\int_0^1 \left(w_1 x_p + w_2 x_p^2 + w_3 x_p^3\right)i_{loc,p,FCP2D}dx_p$$

where $i_{loc,n,FCP2D}$ and the $i_{loc,p,FCP2D}$ denote the interfacial current density in the anode and cathode in FCP2D, respectively. The 5 ordinary differential equations Eqs. (58)−(60), (79), (80) can be used to solve $a_{n,0}$, $a_{n,2}$, $a_{n,3}$, $a_{s,2}$ and $a_{p,3}$.

Next, we construct the 2 equations to solve $b_{n,3}$, $b_{p,3}$ for the electrolyte potential. Substituting the electrolyte potential in Eq. (33) and $w(x) = w_1 x_i + w_2 x_i^2 + w_3 x_i^3$ into Eq. (53), we get

$$\beta\int_0^1 \left(w_1 x_n + w_2 x_n^2 + w_3 x_n^3\right)\left(\frac{(2a_{n,2}+6a_{n,3}x_n)(a_{n,0}+a_{n,1}x_n+a_{n,2}x_n^2+a_{n,3}x_n^3)-(a_{n,1}+2a_{n,2}x_n+3a_{n,3}x_n^2)^2}{(a_{n,0}+a_{n,1}x_n+a_{n,2}x_n^2+a_{n,3}x_n^3)^2}\right)dx_n , \quad (81)$$

$$= w_1\left(b_{n,2}+2b_{n,3}\right)+w_2\left(\frac{2}{3}b_{n,2}+\frac{3}{2}b_{n,3}\right)+w_3\left(\frac{1}{2}b_{n,2}+\frac{6}{5}b_{n,3}\right)+\frac{a_{s,n}L_n^2}{\kappa_{e,n}^{eff}}\int\left(w_1 x_n + w_2 x_n^2 + w_3 x_n^3\right)i_{loc,n}dx_n$$

$$\beta\int_0^1 \left(w_1 x_p + w_2 x_p^2 + w_3 x_p^3\right)\left(\frac{(2a_{p,2}+6a_{p,3}x_p)(a_{p,0}+a_{p,1}x_p+a_{p,2}x_p^2+a_{p,3}x_p^3)-(a_{p,1}+2a_{p,2}x_p+3a_{p,3}x_p^2)^2}{(a_{p,0}+a_{p,1}x_p+a_{p,2}x_p^2+a_{p,3}x_p^3)^2}\right)dx_p , \quad (82)$$

$$= w_1\left(b_{p,2}+2b_{p,3}\right)+w_2\left(\frac{2}{3}b_{p,2}+\frac{3}{2}b_{p,3}\right)+w_3\left(\frac{1}{2}b_{p,2}+\frac{6}{5}b_{p,3}\right)+\frac{a_{s,p}L_p^2}{\kappa_{e,p}^{eff}}\int\left(w_1 x_p + w_2 x_p^2 + w_3 x_p^3\right)i_{loc,p}dx_p$$

Note that in the above four equations, the interfacial current density along the thickness direction is position-dependent and cannot be solved analytically. We first calculate the $m_n$ and $m_p$ by Eqs. (74) and (75), and then obtain the solid potential distribution in the anode and cathode by

$$\phi_{s,n,FCP2D}(x_n) = \frac{2RT}{F}\ln\left(\frac{\sqrt{m_n^2+4}+m_n}{2}\right)+\phi_{e,n}+U_{eq,n} , \quad (83)$$

$$\phi_{s,p,FCP2D}(x_p) = -\frac{2RT}{F}\ln\left(\frac{\sqrt{m_p^2+4}+m_p}{2}\right) + \phi_{e,p} + U_{eq,p}, \tag{84}$$

Next, we calculate the position-dependent over-potential using $\phi_{s,n,FCP2D}(x_n)$ and $\phi_{s,p,FCP2D}(x_p)$ by

$$\eta_{i,FCP2D} = \phi_{s,i,RSPM}(x_i) - \phi_{e,i} - U_{eq,i}, \tag{85}$$

The position-dependent local interfacial current density in the anode and cathode ($i_{loc,n,FCP2D}$ and $i_{loc,p,FCP2D}$) is calculated by

$$i_{loc,i,FCP2D} = i_{0,i,FCP2D}\left[\exp\left(\frac{\alpha F\eta_{i,FCP2D}}{RT}\right) - \exp\left(-\frac{(1-\alpha)F\eta_{i,FCP2D}}{RT}\right)\right], \tag{86}$$

where

$$i_{0,i,FCP2D} = Fk_i\left(c_{s,\max,i} - c_{s,surf,i}\right)^{\alpha}\left(c_{s,surf,i}\right)^{1-\alpha}\left(c_{e,i}\right)^{\alpha}. \tag{87}$$

We substitute the $i_{loc,n,FCP2D}$ and the $i_{loc,p,FCP2D}$ calculated from the previous time step into Eqs (79)−(82) and solve them numerically for the next time step.

The terminal voltage of the battery cell is given

$$V_{t,FCP2D} = \phi_{s,p,FCP2D}(x_p=1) - \phi_{s,n,FCP2D}(x_n=0). \tag{88}$$

2.3. *Simulation set-up*

To compare and evaluate the performance of the proposed reduce-order electrochemical models, we introduce two charging/discharging modes in this research. The first mode uses constant current to firstly charge and then discharge the battery. The constant current charging and discharging C rates include 0.5 C, 1.0 C, 2.0 C, 3.0 C, 4.0 C and 5.0 C. For the constant current charging, the cutoff voltage is set to be 4.2 V. For the constant current discharging, the cutoff voltage is set to be 3.2 V. The second mode is to use random current density profiles to mimic the dynamic charging/discharging driving profiles during the usage of batteries. Three random current density profiles as shown in **Fig. 2a** are used in the study. Specifically, each random current density is generated by the superposition of many sinusoidal waves of different amplitude and frequency (see **Appendix B**). The C rate of the dynamic driving profile 1, profile 2, and profile 3 ranges from -1.4 to 1.2, -4.4 to 3.5, and -5.0 to 4.2, respectively. The simulation time for each dynamic driving profile is 1000 seconds. For the second mode, the initial

SOC ($c_{s,surf,i} / c_{s,max,i}$) of the cathode and anode is set to be 0.5. The equilibrium potential of the electrode materials are shown in **Fig. 2b**. The detailed parameter values used in the simulations are summarized in **Appendix B**. The weights in the shape function, $w_1$, $w_2$, and $w_3$ (see Eq. (51)) are set to be 1.0, -3.0, -2.0 for the RSPM, and 1.0, -1.6, -0.6 for the FCP2D, respectively. These values are selected based on the optimization of the results.

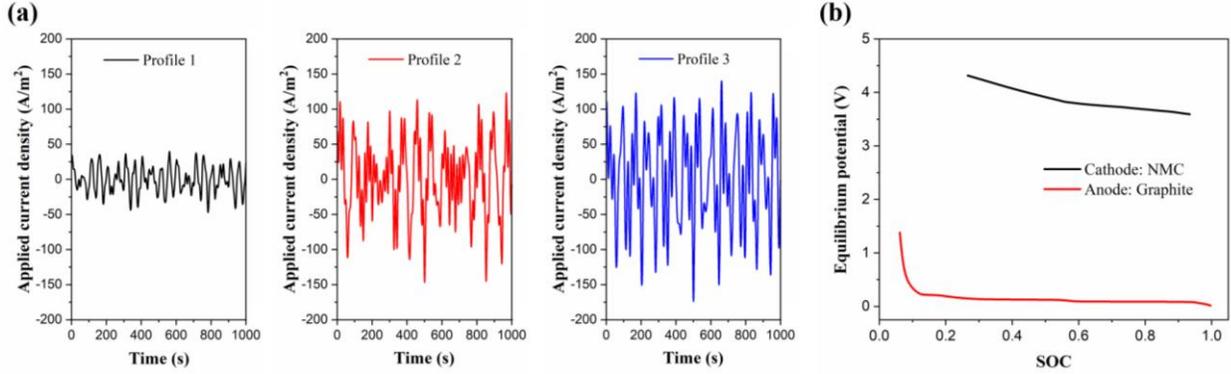

**Fig. 2.** (a) The charging/discharging profiles for the dynamic driving simulations. (b) The equilibrium potential of NMC and graphite electrode materials.

**3. Results and discussions**

To evaluate the accuracy and performance of the developed approach, we use the finite element method (FEM) calculation of the P2D electrochemical model as the benchmark for comparison.

**Fig. 3a** and **3b** show the comparison of terminal voltage of the battery cell calculated by the P2D model, the RSPM, and the FCP2D in the constant current charging/discharging mode. Under lower C-rates (e.g., below 2.5 C), the calculated terminal voltage curves by the RSPM and the FCP2D both agree well with those calculated by the P2D model. As shown in **Fig. 3c**, the voltage error is below 1% compared to the terminal voltage calculated by the P2D model during charging. During discharging, the terminal voltage calculated by the FCP2D and RSPM maintains an error below 1.0% most of the time as shown in **Fig. 3d**. When the C-rate is higher (e.g., above 2.5 C), the FCP2D can still maintain a high accuracy in calculating the terminal voltage, with the voltage error below 2.0%. By contrast, the terminal voltage calculated by the RSPM has less accuracy than the FCP2D when the C-rate is high. The error of the RSPM could be higher than 5.0% at the end of discharging, as shown in **Fig. 3d**. The RSPM is faster than the FCP2D, so the RSPM would be desirable when the C rate is below 2.5 C or when the accuracy requirement is not high.

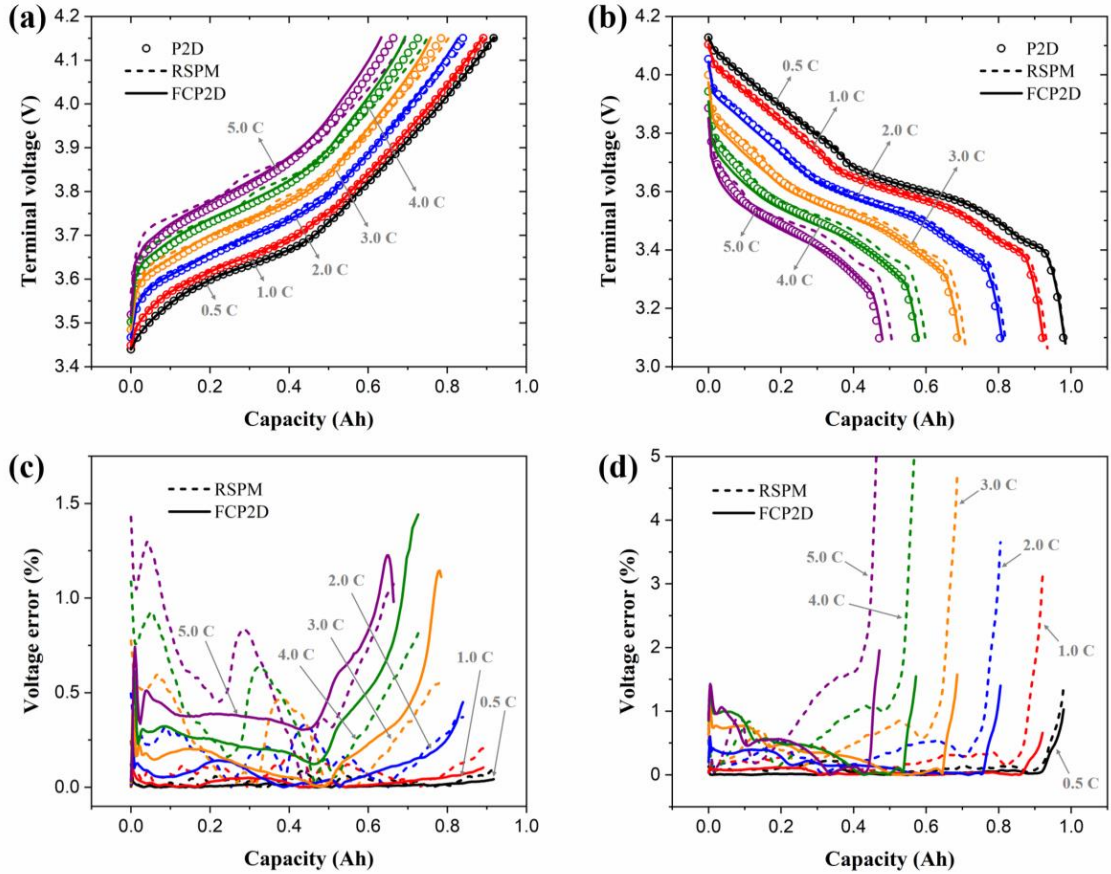

**Fig. 3.** A comparison of terminal voltage calculated by the RSPM, the FCP2D and the P2D model during (a) charging and (b) discharging under various C rates. The calculation error of the RSPM and the FCP2D compared to the P2D model during (c) charging and (d) discharging under various C rates. (The black dots and curves indicate 0.5 C. The red dots and curves indicate 1.0 C. The blue dots and curves indicate 2.0 C. The orange dots and curves indicate 3.0 C. The green dots and curves indicate 4.0 C. The purple dots and curves indicate 5.0 C.)

**Fig. 4** shows a comparison of the lithium ion concentration distribution in the electrolyte ($c_{e,i}$). When the average anode SOC reaches 0.5, **Fig. 4a** and 4**b** show that the lithium ion concentration distribution calculated by the RSPM and the FCP2D both agree well with those calculated by the P2D model under lower C rates. By contrast, under higher C rates, the FCP2D still agree well with the P2D model, while the RSPM shows noticeable difference especially near the current collectors. At the end of charging and discharging, **Fig. 4c** and 4**d** show that the FCP2D agrees well with the P2D model under all C rate scenarios. The RSPM model agrees well with the P2D model under low C rates at the end of the charging process, but shows a larger error than the FCP2D under higher C rates especially near the anode current collector. By contrast, the RSPM agree well with the P2D model under all C rate scenarios at the end of discharging. These results show that the FCP2D can capture the characteristics of lithium ion concentration with a high accuracy at the end of the charging and discharging. The RSPM has a high accuracy in capturing the lithium ion concentration at the end of charging and discharging, especially under low C rates. In the middle of charging or discharging process, the FCP2D can still capture the lithium concentration with a high accuracy, while the RSPM has a relatively high accuracy under low C rates, but the accuracy drops noticeably under high C rates.

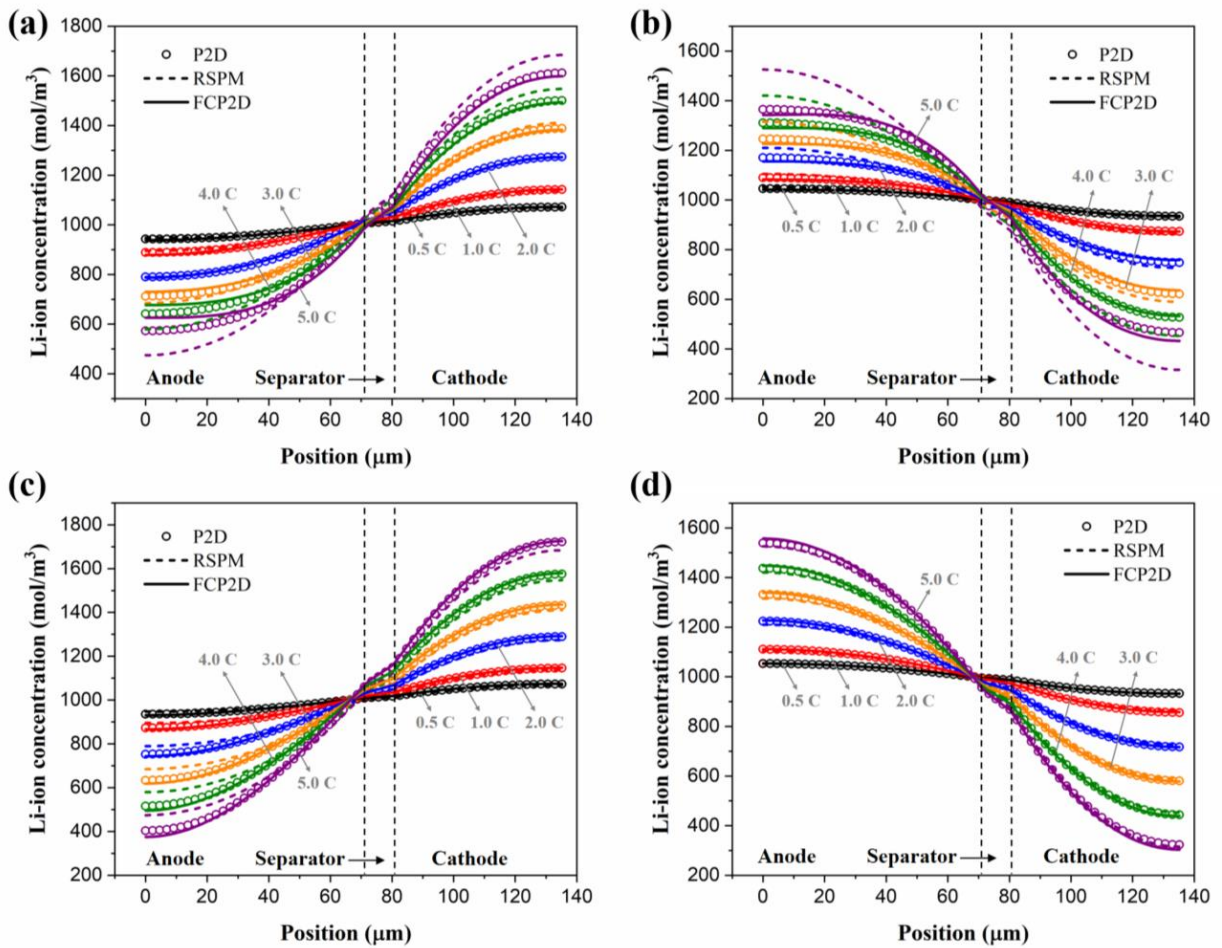

**Fig. 4.** A comparison of the lithium ion concentration distribution in the electrolyte along the electrode thickness direction, calculated by the RSPM, the FCP2D and the P2D model during (a) charging, and (b) discharging under various C rates when the average anode SOC reaches 0.5. A comparison of the lithium ion concentration distribution in the electrolyte along the electrode thickness direction, calculated by the RSPM, the FCP2D and the P2D model at the end of (c)

charging, and (d) discharging. (The black dots and curves indicate 0.5 C. The red dots and curves indicate 1.0 C. The blue dots and curves indicate 2.0 C. The orange dots and curves indicate 3.0 C. The green dots and curves indicate 4.0 C. The purple dots and curves indicate 5.0 C.)

**Fig. 5** shows a comparison of the electrolyte potential distribution ($\phi_{e,i}$). When the average anode SOC reaches 0.5, **Fig. 5a** and **5b** show that the RSPM and the FCP2D both agree well the P2D model under low C rates. Under high C rates, the FCP2D maintains a high accuracy in agreeing with the P2D model, while the RSPM model shows a noticeable deviation. At the end of charging, **Fig. 5c** shows that the electrolyte potential distribution curves calculated by the RSPM and the FCP2D both agree very well with those calculated by the P2D model. At the end of the discharging process, **Fig. 5d** shows that the electrolyte potential distribution curves calculated by the RSPM and the FCP2D are accurate at most locations along the thickness direction of cell. At the location near the cathode current collector, the RSPM is less accurate than the FCP2D, and both show noticeable difference from the electrolyte potential calculated by the P2D model. These results show that in the middle of the charging and discharging process, the FCP2D has a higher accuracy in capturing the electrolyte potential distribution than the RSPM. At the end of charging, both the FCP2D and the RSPM can accurately capture the electrolyte potential distribution. At the end of discharging, the RSPM can accurately capture the electrolyte potential in most locations but will less accuracy than the FCP2D, especially in the regions near the cathode current collector.

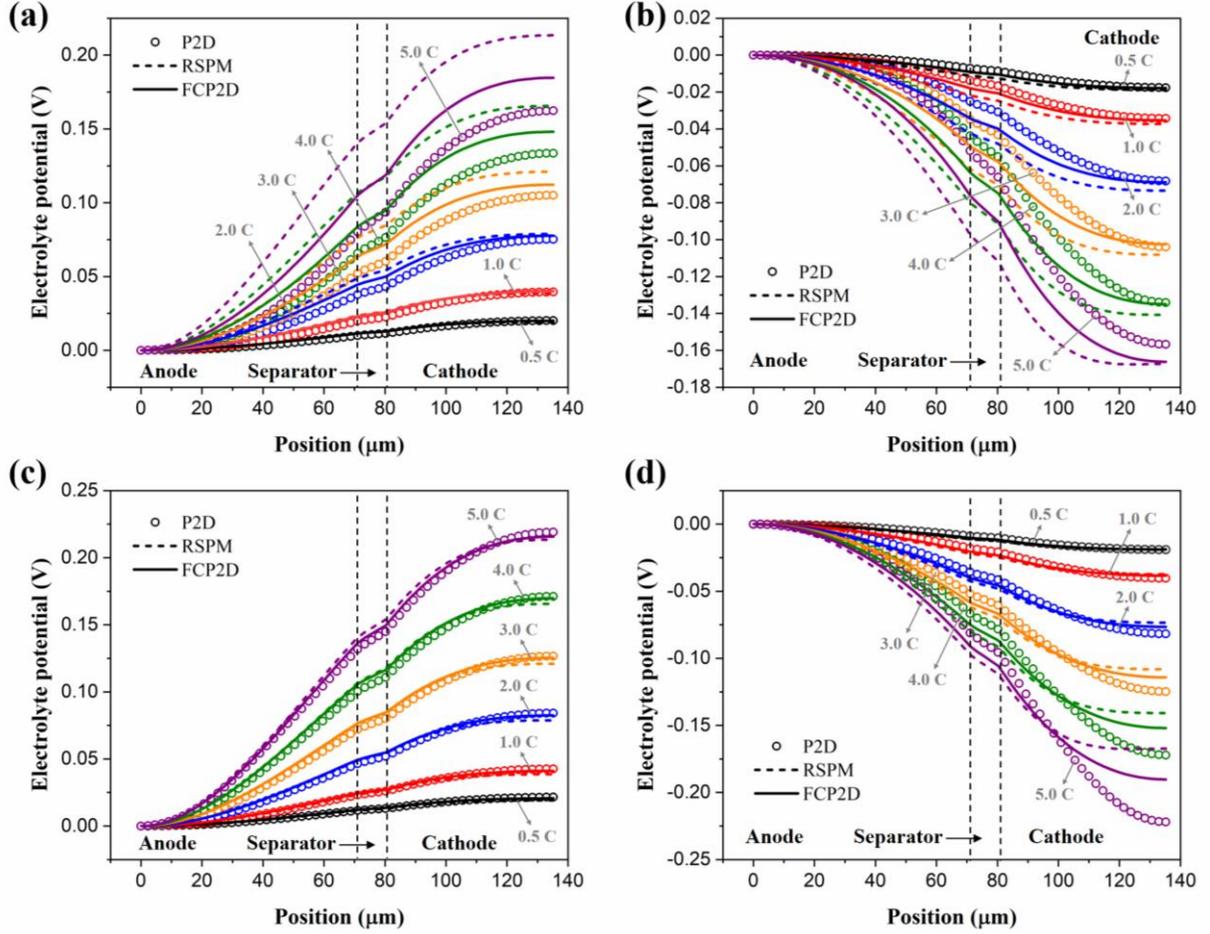

**Fig. 5.** A comparison of the electrolyte potential distribution along the electrode thickness direction, calculated by the RSPM, the FCP2D and the P2D model during (a) charging, and (b) discharging when the average anode SOC reaches 0.5. A comparison of the electrolyte potential distribution along the electrode thickness direction, calculated by the RSPM, the FCP2D and the P2D model at the end of (c) charging, and (d) discharging. (The black dots and curves indicate 0.5 C. The red dots and curves indicate 1.0 C. The blue dots and curves indicate 2.0 C. The orange dots and curves indicate 3.0 C. The green dots and curves indicate 4.0 C. The purple dots and curves indicate 5.0 C.)

To understand the factors that may contribute to the accuracy of the calculated cell voltage, we further analyzed the distribution of lithium concentration on the particle surface ($c_{s,surf,i}$) and the distribution of interfacial current density ($i_{loc,i}$) along the electrode thickness. These results are presented in **Fig. 6**. **Fig. 6a** and **6b** show that in the middle of charging and discharging, the $c_{s,surf,i}$ curves calculated by the FCP2D agree with those calculated by the P2D model under all C rate scenarios. This observation, together with the previous results of the FCP2D for calculating the lithium ion concentration and electrolyte potential distribution, indicate that the FCP2D can capture the electrochemical process accurately under all C-rate scenarios. A notable reason is that the FCP2D can capture the $i_{loc,i}$ distribution along the electrode thickness direction with high accuracy under all C rates considered (see **Fig. 6c** and **6d**).

By contrast, the $i_{loc,i}$ in the RSPM is uniform along the electrode thickness direction because of the model assumption. This treatment simplifies the calculation while providing high accuracy at low C rates, but causes reduced accuracy under high C rates (e.g., above 2.5 C) where the model assumption of uniform $i_{loc,i}$ deviates from the real situation. As a result, the terminal voltage calculated by the RSPM under high C rates is less accurate than that calculated by the FCP2D. At lower C rates, the $i_{loc,i}$ distribution calculated by the RSPM is almost the same as the average of the $i_{loc,i}$ calculated by the P2D model. The RSPM can accurately capture the electrochemical characteristics (e.g., lithium ion concentration in the electrolyte, electrolyte potential, and lithium concentration on the surface of the particles) at lower C rates, and thus provides a high accuracy in the calculated terminal voltage.

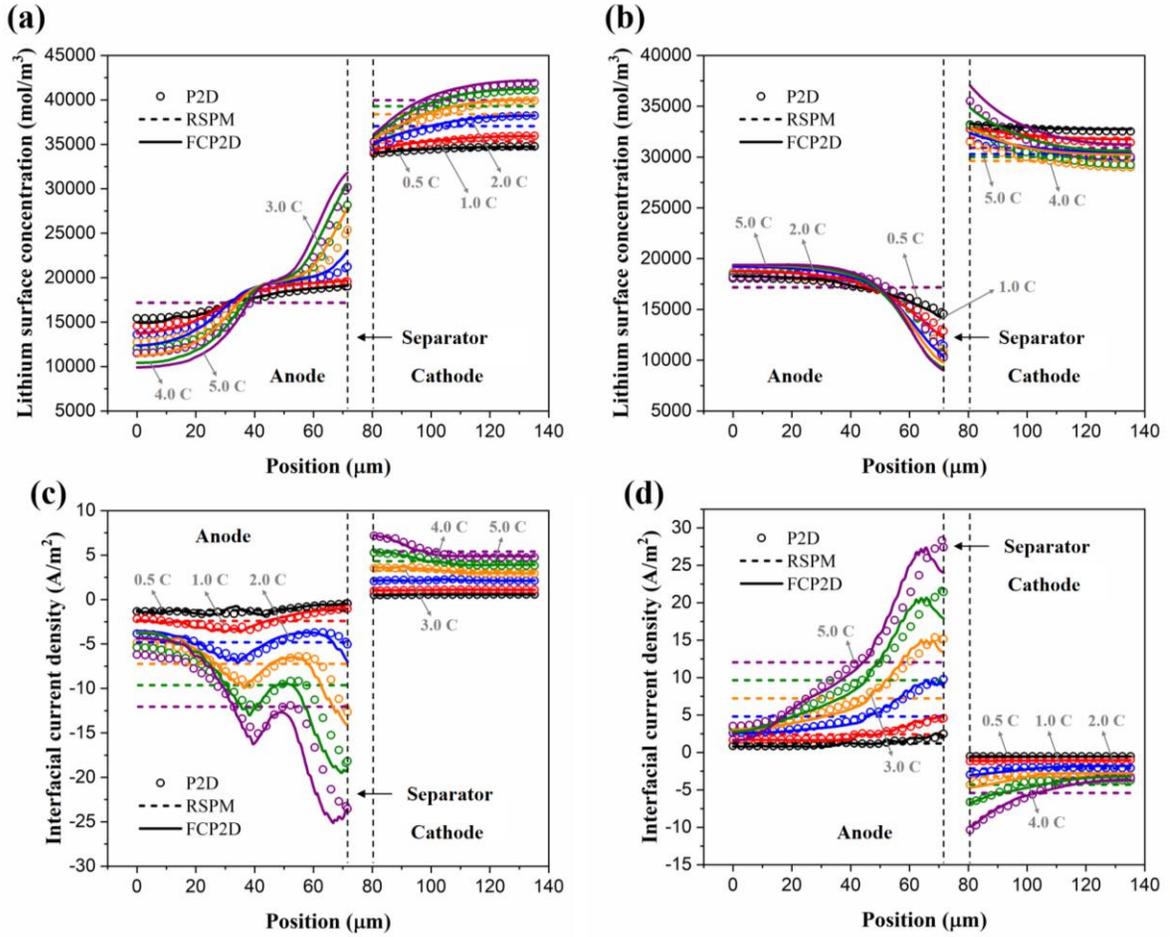

**Fig. 6.** A comparison of the particle surface lithium concentration distribution along the electrode thickness direction, calculated by the RSPM, the FCP2D and the P2D model during (a) charging, and (b) discharging under various C rates when the average anode SOC reaches 0.5. A comparison of the interfacial current density distribution along the electrode thickness direction, calculated by the RSPM, the FCP2D and the P2D model during (a) charging, and (b) discharging under various C rates when the average anode SOC reaches 0.5. (The black dots and curves indicate 0.5 C. The red dots and curves indicate 1.0 C. The blue dots and curves indicate 2.0 C. The orange dots and curves indicate 3.0 C. The green dots and curves indicate 4.0 C. The purple dots and curves indicate 5.0 C.)

**Fig. 7** compares the time consumption of the models in completing the calculation of one charging and discharging cycle. The time step size of the RSPM, the FCP2D, and the P2D model are set to be the same, which is (10/C rate) second for each C rate. For instance, the time step for 1 C charging and discharging is 10 s. We can observe that for all three models, the time consumption decreases monotonically with the C rate. For each C rate, the P2D model is the slowest, leading to the longest time consumption. The FCP2D is significantly faster, while the RSPM is the fastest with the least time consumption. The time consumption of RSPM is only 3.59%, 3.79%, 4.19%, 3.97%, 3.57%, and 3.05% of that of the P2D model under the C rate of 0.5, 1.0, 2.0, 3.0, 4.0, and 5.0, respectively. The time consumption of FCP2D is 12.35%, 14.22%, 12.57%, 13.24%, 13.57%, and 12.21% of that of the P2D model, respectively. These results show that the RSPM is much faster than the FCP2D. However, when the C rate is large (e.g., above 2.5 C), the RSPM is less accurate than the FCP2D. Thus, a trade-off between accuracy and efficiency exists when selecting between these two models.

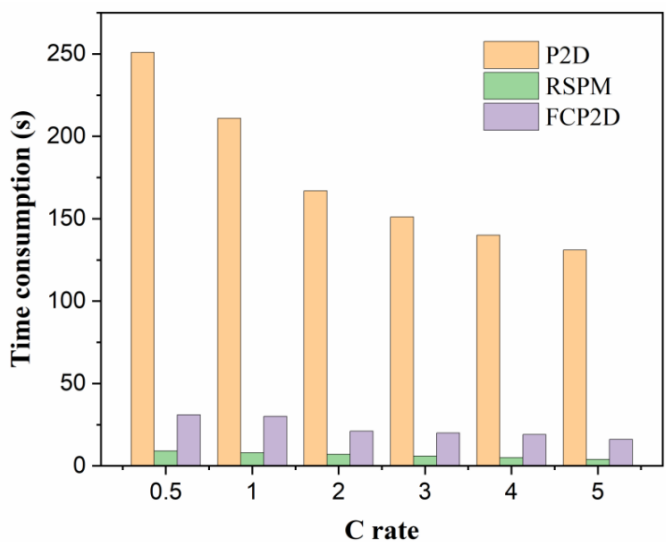

**Fig. 7.** The time consumption of the RSPM, the FCP2D and the P2D model for simulating one complete constant current charging and discharging cycle under various C rates.

The accuracy of the RSPM and the FCP2D are further evaluated under the dynamic driving profiles (**Fig. 2a**) composed of random charging/discharging current densities and frequency. The calculated terminal voltage curves are shown in **Fig. 8**. It can be seen from **Fig. 8a**-**8c** that for all the three dynamic driving profiles, the terminal voltage curves calculated by the RSPM and the FCP2D both agree well with those calculated by the P2D model. The average voltage error (averaged over the 1000 s simulate time) of the RSPM s is 0.054%, 0.148%, and 0.190% in the dynamic driving profile 1, dynamic driving profile 2, and dynamic driving profile 3, respectively. The average voltage error of the FCP2D is 0.049%, 0.130%, and 0.174%, respectively. For the dynamic driving profile 1, **Fig. 8d** shows that the voltage error of the RSPM is mostly below 0.120% (indicated by the blue dash line). There are few moments when the voltage error reaches around 0.160%, and rarely up to 0.180%. The voltage error of the FCP2D is mostly below 0.100% (indicated by the red dash line). There are few moments when the transient voltage error reaches around 0.14%, and rarely up to 0.168%. For the dynamic driving profile 2, **Fig. 8e** shows that the voltage error of the RSPM is mostly below 0.400% (indicated by the blue dash line), there is only one moment when the voltage error reaches 0.900%. For the FCP2D, the voltage error is mostly below 0.300% (indicated by the red dash line). There are few moments when the voltage error reaches around 0.350%, and rarely up to 0.460%. This peak is significantly lower than the maximum voltage error of the RSPM. For the dynamic driving profile 3, **Fig. 8f** shows that the voltage error of the RSPM is mostly below 0.450% (indicated by the blue dash line). There are few moments when the voltage error reaches around 0.500%, and rarely up to 0.600%. For the FCP2D, the voltage error is mostly below 0.375% (as indicated by the red dash line). There are few moments when the voltage error reaches around 0.400%, and rarely up to 0.570%.

These results indicate that the RSPM and the FCP2Ds can both provide a high accuracy in simulating the battery terminal voltage under dynamic driving scenarios. The FCP2D has a higher average accuracy (lower average voltage error) than the RSPM. The voltage error of the RSPM and the FCP2Ds both increases when the C rate range of the random current density profile is larger. The FCP2D is more accurate than the RSPM for all the dynamic driving profiles, as reflected by the smaller voltage error in most time.

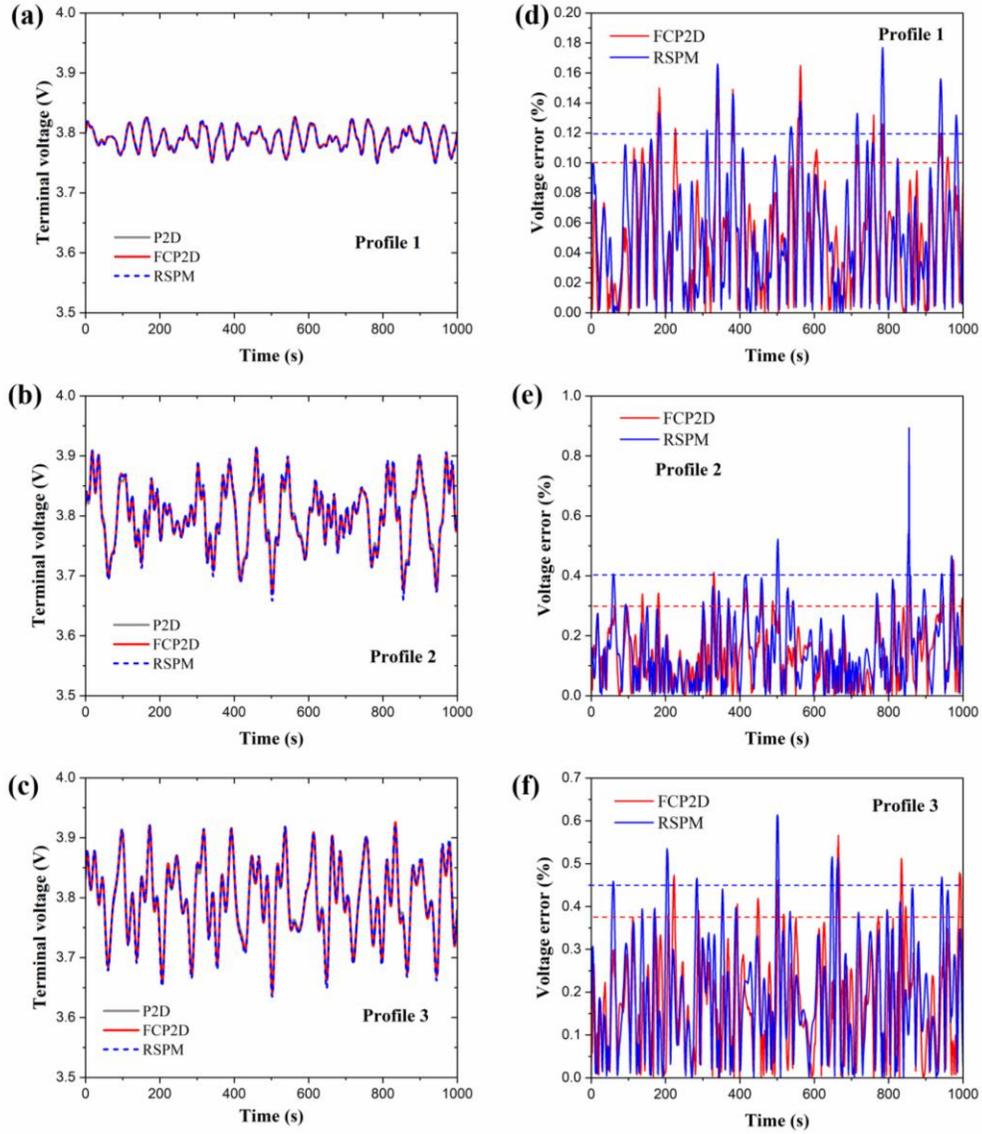

**Fig. 8.** A comparison of the terminal voltage curves calculated by the RSPM, the FCP2D and the P2D model in (a) dynamic driving profile 1, (b) dynamic driving profile 2, and (c) dynamic driving profile 3. The error of the terminal voltage calculated by the RSPM and the FCP2D in (d) dynamic driving profile 1, (e) dynamic driving profile 2, and (f) dynamic driving profile 3.

## 4. Conclusions

This work proposes two novel reduced-order electrochemical models, revised single particle model (RSPM) and fast calculation P2D model (FCP2D), to accelerate battery simulation while maintaining high accuracy under various charging/discharging current density profiles. Specifically, the RSPM is constructed with the assumption of two single particles representing the two electrodes for the electrochemical activities happening on the particle surfaces. The FCP2D is constructed with multiple particles contained in each electrode at each location. We model the electrolyte potential distribution and lithium-ion concentration distribution along the battery thickness direction in both RSPM and FCP2D as polynomial functions. We propose a method of shape functions to solve the unknown polynomial parameters, and obtain their weights by optimization.

Results show that the proposed RSPM and FCP2D can generate the battery terminal voltage and electrochemical parameters, including the distribution of lithium-ion concentration, electrolyte potential, interfacial current density, and lithium concentration on the surface of particles with high accuracy and speed. They are both much faster than the P2D electrochemical model. The FCP2D has a higher accuracy than the RSPM, especially at higher C rates, while the RSPM is faster. Both of them can predict the battery terminal voltage accurately under random charging/discharging profiles.

**Acknowledgement**

The authors gratefully acknowledge the support by LG Energy Solution.

**Appendix A: Derivation of the electrolyte potential in the separator region**

In the separator region, we have

$$\frac{\partial}{\partial x_s}\left(\frac{\partial \phi_{e,s}}{\partial x_s} - \beta \frac{\partial \ln c_{e,s}}{\partial x_s}\right) = 0, \tag{A1}$$

Integrating both sides, we have

$$\frac{\partial \phi_{e,s}}{\partial x_s} - \beta \frac{\partial \ln c_{e,s}}{\partial x_s} + cons1 = 0, \tag{A2}$$

$$\phi_{e,s} - \beta \ln c_{e,s} + cons1 \cdot x_s + cons2 = 0. \tag{A3}$$

The continuity conditions are

$$\phi_{e,n}\big|_{x_n=1} = \phi_{e,s}\big|_{x_s=0}, \tag{A4}$$

$$\phi_{e,s}\big|_{x_s=1} = \phi_{e,p}\big|_{x_p=0}, \tag{A5}$$

$$-\frac{\kappa_{e,s}^{eff}}{L_s}\left(\frac{\partial \phi_{e,s}}{\partial x_s} - \beta \frac{\partial \ln c_{e,s}}{\partial x_s}\right)\bigg|_{x_s=0} = i_{app}, \tag{A6}$$

$$-\frac{\kappa_{e,s}^{eff}}{L_s}\left(\frac{\partial \phi_{e,s}}{\partial x_s} - \beta \frac{\partial \ln c_{e,s}}{\partial x_s}\right)\bigg|_{x_s=1} = i_{app}, \tag{A7}$$

After substitution of Eq. (A2), the conditions in Eqs. (A6) and (A7) both give $cons1 = L_s i_{app} / \kappa_{e,s}^{eff}$.

Substituting Eq. (A3), $c_{e,s} = a_{s,0} + a_{s,1}x_s + a_{s,2}x_s^2$ and $\phi_{e,n} = b_{n,0} + b_{n,1}x_n + b_{n,2}x_n^2 + b_{n,3}x_n^3$ into the condition in Eq. (A4) and noting that $c_{e,s}\big|_{x_s=0} = a_{s,0}$ and $\phi_{e,n}\big|_{x_n=1} = b_{n,2} + b_{n,3}$ (Eqs. (41) and (42) show $b_{n,0} = 0$ and $b_{n,1} = 0$), we get $cons2 = \beta \ln(a_{s,0}) - b_{n,2} - b_{n,3}$. The electrolyte potential in the separator region is then given by

$$\phi_{e,s} = \beta \ln\left(a_{s,0} + a_{s,1}x_s + a_{s,2}x_s^2\right) - \frac{L_s i_{app}}{\kappa_{e,s}^{eff}} x_s + b_{n,2} + b_{n,3} - \beta \ln(a_{s,0}). \tag{A8}$$

The condition in Eq. (A5), together with $\phi_{e,p} = b_{p,0} + b_{p,1}x_p + b_{p,2}x_p^2 + b_{p,3}x_p^3$ and the potential expression of Eq. (A8), gives a relation

$$b_{p,0} = \beta \ln\left(\frac{a_{s,0} + a_{s,1} + a_{s,2}}{a_{s,0}}\right) - \frac{L_s i_{app}}{\kappa_{e,s}^{eff}} + b_{n,2} + b_{n,3}. \tag{A9}$$

Substituting the expression of $b_{n,2}$ in Eq. (43) to Eq. (A9) gives

$$b_{p,0} = \beta \ln\left(\frac{a_{s,0} + a_{s,1} + a_{s,2}}{a_{s,0}}\right) - \left(\frac{L_s}{\kappa_{e,s}^{eff}} + \frac{L_n}{2\kappa_{e,n}^{eff}}\right)i_{app} + \frac{\beta}{2}\frac{2a_{n,2} + 3a_{n,3}}{a_{n,0} + a_{n,2} + a_{n,3}} - \frac{b_{n,3}}{2}. \tag{A10}$$

**Appendix B: Parameters used in this work**

**Table B1.** Values of electrochemical parameters used in simulations.

| Parameter | Symbol | Value |
|---|---|---|
| $c_{e0}$ | Initial lithium-ion concentration | 1000 mol m$^{-3}$ |
| $c_{s,\max,n}$ | Maximum lithium-ion concentration in the anode particle | 34347 mol m$^{-3}$ |
| $c_{s,\max,p}$ | Maximum lithium-ion concentration in the cathode particle | 54789 mol m$^{-3}$ |
| $D_{e0}$ | Lithium ion diffusion coefficient in bulk electrolyte | $4\times10^{-10}$ m$^2$ s$^{-1}$ |
| $D_{s,n}$ | Lithium diffusion coefficient in the solid phase of anode | $2.93\times10^{-14}$ m$^2$ s$^{-1}$ |
| $D_{s,p}$ | Lithium diffusion coefficient in the solid-phase of cathode | $1.00\times10^{-12}$ m$^2$ s$^{-1}$ |
| $F$ | Faraday constant | 96485 C mol$^{-1}$ |
| $k_n$ | Reaction rate constant of anode | $3.08\times10^{-10}$ m s$^{-1}$ |
| $k_p$ | Reaction rate constant of cathode | $1.30\times10^{-10}$ m s$^{-1}$ |
| $L_n$ | Thickness of anode | 71.60 μm |
| $L_p$ | Thickness of cathode | 54.62 μm |
| $L_s$ | Thickness of separator | 9.00 μm |
| $r_{p,n}$ | Radius of the anode particle | 10.00 μm |
| $r_{p,p}$ | Radius of the cathode particle | 3.75 μm |
| $R$ | Ideal gas constant | 8.3145 J mol$^{-1}$ K$^{-1}$ |
| $t_+^0$ | Lithium-ion transference number | 0.363 |
| $\alpha$ | Anodic charge transfer coefficient | 0.5 |
| $\varepsilon_{e,n}$ | Electrolyte volume fraction in anode | 0.315 |
| $\varepsilon_{e,p}$ | Electrolyte volume fraction in cathode | 0.265 |
| $\varepsilon_{e,s}$ | Electrolyte volume fraction in separator | 0.450 |
| $\varepsilon_{s,n}$ | Solid phase material volume fraction in anode | 0.585 |
| $\varepsilon_{s,p}$ | Solid phase material volume fraction in cathode | 0.635 |
| $\kappa_{e0}$ | Bulk electrolyte conductivity | 0.95 S m$^{-1}$ |
| $\sigma_{s0,n}$ | Bulk solid phase conductivity of anode | 50 S m$^{-1}$ |
| $\sigma_{s0,p}$ | Bulk solid phase conductivity of cathode | 13.75 S m$^{-1}$ |

The formula for generating the dynamic driving profile is

$$i_{app} = A_1 \sin(\omega_1 t) + A_2 \cos(\omega_2 t) + A_3 \sin(\omega_3 t) \\ + A_4 \cos(\omega_4 t) + A_5 \sin(\omega_5 t) + A_6 \cos(\omega_6 t)$$

(B1)

The parameters are shown in Table B2.

Table B2. Values of current density profile parameters

| Parameter | Profile 1 | Profile 2 | Profile 3 |
|---|---|---|---|
| $A_1$ | 0.600 | 1.200 | 2.000 |
| $A_2$ | 0.205 | 0.48 | 0.800 |
| $A_3$ | 0.125 | 0.336 | 0.560 |
| $A_4$ | 0.360 | 0.864 | 1.440 |
| $A_5$ | 0.070 | 1.368 | 0.280 |
| $A_6$ | 0.180 | 0.504 | 0.720 |
| $\omega_1$ | 0.126 | 0.086 | 0.056 |
| $\omega_2$ | 0.043 | 0.143 | 0.163 |
| $\omega_3$ | 0.311 | 0.211 | 0.234 |
| $\omega_4$ | 0.157 | 0.357 | 0.257 |
| $\omega_5$ | 0.472 | 0.072 | 0.172 |
| $\omega_6$ | 0.325 | 0.395 | 0.295 |